\documentclass[aps,prd,twocolumn,superscripts,showkeys,nofootinbib]{revtex4-2}

\usepackage{amsmath,amssymb,amsfonts}
\usepackage{graphicx}
\usepackage{dcolumn}
\usepackage{bm}
\usepackage{hyperref}
\usepackage{xcolor}
\usepackage{booktabs}

\newcommand{\Msun}{M_{\odot}}
\newcommand{\muB}{\mu_{B}}
\newcommand{\alphaGB}{\alpha}
\newcommand{\cs}{c_{s}}
\newcommand{\Mmax}{M_{\max}}
\newcommand{\rhoC}{\rho_{c}}

\begin{document}

\title{Quark Stars in Regularized 4D Einstein--Gauss--Bonnet Gravity:
       A Perturbative QCD Equation of State}

\author{Anirudh Pradhan}
\email{pradhan.anirudh@gmail.com}
\affiliation{Centre for Cosmology, Astrophysics and Space Science,
GLA University, Mathura-281\,406, Uttar Pradesh, India}

\author{Safiqul Islam}
\email{corresponding author: sislam@kfu.edu.sa }
\affiliation{Department of Mathematics and Statistics, College of Science, King Faisal University, P.O. Box 400, Al Ahsa 31982, Saudi Arabia}

\author{Safyan Mukhtar}
\email{smahmad@kfu.edu.sa}
\affiliation{Department of Mathematics and Statistics, College of Science, King Faisal University, P.O. Box 400, Al Ahsa 31982, Saudi Arabia}

\author{Santosh Kumar Dixit}
\email{skdixit@ptn.amity.edu}
\affiliation{Amity School of Engineering and Technology, Amity University Patna, Patna-801503, India}

\date{\today}

\begin{abstract}
We investigate the equilibrium structure and stability of selfbound
quark stars in the framework of regularized four-dimensional
Einstein--Gauss--Bonnet (4DEGB) gravity, employing the perturbative
QCD equation of state of Fraga, Kurkela and Vuorinen~\cite{Fraga:2014interacting}.
The equation of state, parameterised by a single
renormalization-scale parameter $X$, contains no effective bag
constant and defines the stellar surface entirely through the
vanishing of the quark-matter pressure. We solve the modified
Tolman--Oppenheimer--Volkoff equations derived from the scalar--tensor
formulation of 4DEGB gravity for $X \in \{3, 4\}$ and Gauss--Bonnet
coupling $\alpha \in \{0, 1, 10\}\,\mathrm{km}^2$. For the soft EOS
($X = 3$), the maximum mass increases from $2.0431\,M_\odot$ (GR) to
$2.5013\,M_\odot$ at $\alpha = 10\,\mathrm{km}^2$, with only the
latter entering the PSR~J0952$-$0607 mass band. For the stiff EOS
($X = 4$), the GR baseline already yields $3.0415\,M_\odot$, and all
configurations exceed both the GW190814 secondary mass and the
PSR~J0952$-$0607 constraint. The compactness $C = M/R$ at maximum
mass ranges from $0.2531$ to $0.3123$ across all configurations,
remaining well below the Buchdahl bound throughout. Radial profiles
of the squared speed of sound confirm that $c_s^2 < 1/3$ holds
pointwise throughout the stellar interior for all parameter choices,
establishing that the quark matter remains sub-conformal and causal
inside the maximum-mass star. These results demonstrate that higher-curvature corrections in 4DEGB
gravity systematically enhance the maximum supported mass of perturbative
QCD quark stars, with the stiff ($X = 4$) branch already exceeding the
most massive compact objects currently observed even at the GR level,
while the soft ($X = 3$) branch requires $\alphaGB \sim 10\,\mathrm{km}^2$
to approach those thresholds.
\end{abstract}


\maketitle

\section{Introduction}
\label{sec:intro}

Compact astrophysical objects such as neutron stars and quark stars
occupy a central position in modern high-energy astrophysics, as they
provide natural laboratories for exploring the behaviour of matter and
gravity under extreme conditions~\cite{Lattimer:2004pg,Ozel:2016oaf}.
Characterised by ultra-high central densities, enormous pressures, and
strong spacetime curvature, these objects probe physical regimes that
are inaccessible to terrestrial experiments. As a result, global stellar
observables---most notably the mass, radius, and
compactness---are highly sensitive to both the microphysical properties
of dense matter and the underlying gravitational
theory~\cite{Lattimer:2006xb}. Recent advances in observational
astrophysics, including precise pulsar mass measurements from Shapiro
delay~\cite{Demorest:2010twosolarmass,Cromartie:2019relativistic,
Fonseca:2021refined}, X-ray timing observations by
NICER~\cite{Riley:2019nicer,Miller:2019psr,Riley:2021nicer,
Miller:2021radius,Choudhury:2024nicer}, and multimessenger detections of
compact binary
mergers~\cite{Abbott:2017gw170817,Abbott:2018gw170817,Abbott:2020gw190814},
have significantly improved our ability to confront theoretical models
with empirical data, thereby opening new avenues for testing fundamental
physics in the strong-field regime~\cite{Berti:2015itd,Psaltis:2008bb,
Yagi:2013awa,Yunes:2013dva}.

Despite the remarkable empirical success of general relativity (GR) in
describing gravitational phenomena from solar-system scales to
strong-field observations involving compact objects and gravitational
waves~\cite{Will:2014kxa}, several fundamental and observational
issues---including the late-time acceleration of the Universe, the
presence of dark matter, and the quest for a consistent quantum theory
of gravity---motivate the exploration of alternative theories of
gravity~\cite{Sotiriou:2008rp,DeFelice:2010aj,Clifton:2011jh}. Among
these extensions, higher-curvature modifications have attracted
sustained interest, with Einstein--Gauss--Bonnet (EGB) gravity standing
out due to its special Lovelock structure, which guarantees second-order
field equations and avoids Ostrogradsky
instabilities~\cite{Lovelock:1971yv,Charmousis:2008kc}. Although the
Gauss--Bonnet term is topological in four spacetime dimensions and does
not contribute dynamically in standard GR, its role as a leading
higher-curvature correction has motivated extensive investigations in
higher-dimensional settings and
beyond~\cite{Boulware:1985wk,Aranguiz:2015voa,Rubiera-Garcia:2015yga,
Giacomini:2015dwa}.

Recent work by Glavan and Lin~\cite{Glavan:2019inb} introduced a novel
prescription in which the Gauss--Bonnet coupling constant $\alpha$ is
rescaled as $\lim_{D\to4}(D-4)\alpha \to \alpha$, leading to what is now
known as the four-dimensional Einstein--Gauss--Bonnet (4DEGB) theory. A
number of criticisms were subsequently raised regarding the
well-definedness of this
procedure~\cite{Gurses:2020ofy,Ai:2020peo,Shu:2020cjw}, which were
addressed by demonstrating that the $D\to4$ limit can be taken
consistently at the level of the gravitational
action~\cite{Hennigar:2020lsl,Fernandes:2020rpa,Mann:1992ar}, yielding
a four-dimensional scalar--tensor theory of the Horndeski
class~\cite{Lu:2020iav}. The resulting theory has been shown to be an
interesting phenomenological competitor to
GR~\cite{Clifton:2020xhc,Fernandes:2022zrq,Zanoletti:2023ori,
Charmousis:2021npl}, and a wide range of physically interesting
solutions have been obtained, including black
holes~\cite{Ghosh:2020syx,Konoplya:2020bxa,Konoplya:2020juj,
Hosseinimansoori:2020yfj,Yang:2020jno,PhysRevD.101.104018}, charged
black holes~\cite{Fernandes:2020nbq,Zhang:2020sjh}, optical and
dynamical properties~\cite{Islam:2020xmy,Jin:2020emq,Guo:2020zmf,
Abdujabbarov:2020jla,Rayimbaev:2020lmz,Zeng:2020dco,Zubair:2023cep,
Rayimbaev:2022znx}, and traversable wormhole
configurations~\cite{Jusufi:2020yus,Jusufi:2020qyw}.

In astrophysical contexts, compact stars provide particularly sensitive
laboratories in which even small deviations from GR can lead to
observable modifications of stellar structure and stability. Since the
formulation of the regularized 4DEGB theory, there has been considerable
interest in its predictions for compact stellar configurations. The
structure of relativistic stars has been studied by Doneva and
Yazadjiev~\cite{Doneva:2020ped} and subsequently by Saavedra et
al.~\cite{Saavedra:2024fzy}, while the constraints from astrophysical
observations were analyzed by Charmousis et
al.~\cite{Charmousis:2021npl}. In the context of quark stars, one of the
present authors and collaborators~\cite{Banerjee:2020stc} investigated
strange quark stars using the MIT bag model equation of state, and in a
subsequent work~\cite{Banerjee:2020yhu} extended this analysis to an
interacting quark matter equation of state characterised by an effective
bag constant and an interaction parameter. Color-flavor locked (CFL)
quark stars in 4DEGB gravity were explored in
Ref.~\cite{Banerjee:2020dad}. Electrically charged quark stars in this
framework were studied by Pretel et al.~\cite{Pretel:2021czp}, and more
recently a unified interacting quark matter equation of state was
employed by Gammon, Rourke and Mann~\cite{Gammon:2023uss}, who also
identified the possibility of extreme compact objects (ECOs) whose radii
lie below the GR Buchdahl bound. White dwarfs in the regularized 4DEGB
theory have been studied in Ref.~\cite{Pretel:2025roz}, while broader
classes of stellar configurations including anisotropic stars and
polytropes have been considered in
Refs.~\cite{Hansraj:2024qbe,Bordbar:2024yai,NewtonSingh:2022rfo,
SINGH2020100730}. A QCD-based equation of state with tidal
deformability was recently investigated by Tangphati et
al.~\cite{Tangphati:2025yih}, and dark-matter--dark-energy admixed
compact stars in the same framework have been studied in
Ref.~\cite{Banerjee:2020stc}. Together these studies demonstrate that
higher-curvature corrections can significantly influence the equilibrium
structure, stability, and observable properties of compact stars.

The equation of state (EOS) of dense quark matter plays an equally
important role in determining the observable properties of quark stars.
A particularly well-motivated description of cold, dense quark matter is
provided by perturbative quantum chromodynamics (pQCD). Early studies of
the QCD equation of state at zero temperature and high density were
carried out by Freedman and McLerran~\cite{Freedman:1977fermions,
Freedman:1978quark} and by Baluni~\cite{Baluni:1978nonabelian}, and
these have been substantially refined through successive higher-order
calculations~\cite{Toimela:1985perturbative,Fraga:2001small,
Fraga:2005role,Kurkela:2010cold,Ghisoiu:2017highorder,
Gorda:2018nexttonexttonexttoleading,Gorda:2021cold}. In a landmark
paper, Fraga, Kurkela and Vuorinen~\cite{Fraga:2014interacting}
cast the three-loop result of Kurkela, Romatschke and
Vuorinen~\cite{Kurkela:2010cold} into a compact pocket formula for
the pressure of cold, beta-equilibrated, charge-neutral three-flavor
quark matter. For brevity, we hereafter refer to this pocket formula
as FKV[$X$]; it encapsulates the perturbative QCD corrections through a
single dimensionless renormalization-scale parameter $X \in [1,4]$,
with $X = 3$ and $X = 4$ representing two values in the upper part of
this range that together span the theoretical uncertainty band of the
calculation. A crucial feature of
the FKV[$X$] EOS is that it contains no effective bag constant and no
additional phenomenological parameters beyond $X$; the surface of the
star is defined entirely by the condition that the pressure vanishes at
the quark-matter surface chemical potential. The implications of this
EOS for compact star structure and for neutron star constraints from QCD
have been analyzed extensively in the
literature~\cite{Kurkela:2014constraining,S.Fraga:2016neutron,
Annala:2018gravitationalwave}.

Motivated by these developments, the present work investigates the
structure and stability of selfbound quark stars in the framework of
regularized 4DEGB gravity, employing the FKV[$X$] equation of
state. By scanning the Gauss--Bonnet coupling parameter $\alpha \in
\{0, 1, 10\}\,\mathrm{km}^2$ and the renormalization-scale parameter
$X \in \{3, 4\}$, we compute the mass--radius relation, the
mass--central-density sequence, the compactness, and the radial profile
of the speed of sound for each configuration. We compare our results
against current observational constraints, including the secondary
component of the binary merger GW190814~\cite{Abbott:2020gw190814},
identified as a compact object with mass $2.59^{+0.08}_{-0.09}\,\Msun$,
and the black widow pulsar PSR~J0952$-$0607~\cite{Romani:2022psr} with
mass $2.35 \pm 0.17\,\Msun$. To the best of our knowledge, this is the
first systematic study of quark stars in regularized 4DEGB gravity
employing the FKV[$X$] pocket formula, which contains
no bag constant and uses the renormalization-scale parameter $X$ as its
sole free input. This approach complements and extends the existing
landscape of quark star studies in 4DEGB gravity by providing a
bag-constant-free, single-parameter pQCD EOS anchored entirely in
perturbative QCD.

The paper is organized as follows. In Sec.~\ref{sec:theory}, we
summarise the gravitational field equations of regularized 4DEGB gravity
and derive the modified TOV equations governing stellar equilibrium.
Section~\ref{sec:eos} describes the FKV[$X$] equation of state,
its thermodynamic properties, and the causality and conformality checks.
Section~\ref{sec:results} presents our numerical results, including the
mass--radius relations, compactness, and the radial sound-speed profiles.
Section~\ref{sec:conclusions} summarises the main findings and outlines
future directions.

\section{Theoretical Framework}
\label{sec:theory}

\subsection{4D Einstein--Gauss--Bonnet Gravity}
\label{subsec:action}

The regularized four-dimensional Einstein--Gauss--Bonnet (4DEGB) theory
is defined by augmenting the Einstein--Hilbert action with a
Gauss--Bonnet term coupled to a scalar field $\phi$, leading to the
scalar--tensor action~\cite{Hennigar:2020lsl,Fernandes:2020rpa}
\begin{widetext}
\begin{equation}
  S = \frac{1}{2\kappa}\int d^4x\sqrt{-g}
      \Bigl[R - 2\Lambda + \alpha\bigl(\phi\,\mathcal{G}
      + 4G^{\mu\nu}\nabla_\mu\phi\nabla_\nu\phi
      - 4(\nabla\phi)^2\Box\phi
      + 2(\nabla\phi)^4\bigr)\Bigr] + S_m,
  \label{eq:action}
\end{equation}
\end{widetext}
where $\kappa = 8\pi$ in natural units ($\hbar = G = c = 1$), $\alpha$
denotes the Gauss--Bonnet coupling constant with dimensions of length
squared, $\phi$ is a dimensionless scalar field, and $S_m$ represents
the matter action. The Gauss--Bonnet invariant appearing in the action
is defined as
\begin{equation}
  \mathcal{G} = R^{\mu\nu\rho\sigma}R_{\mu\nu\rho\sigma}
              - 4R^{\mu\nu}R_{\mu\nu} + R^2.
  \label{eq:GB_invariant}
\end{equation}
An important structural property of the action~\eqref{eq:action} is
its invariance under constant shifts of the scalar field,
\begin{equation}
  \phi \to \phi + C,
  \label{eq:shift_symmetry}
\end{equation}
where $C$ is an arbitrary constant, reflecting the shift symmetry
characteristic of this scalar--tensor formulation.

Variation of the action with respect to the scalar field $\phi$ yields
the scalar field equation of motion
\begin{widetext}
\begin{equation}
  \mathcal{G} - 8G^{\mu\nu}\nabla_\mu\nabla_\nu\phi
  - 8R^{\mu\nu}\nabla_\mu\phi\nabla_\nu\phi
  + 8(\Box\phi)^2 - 8(\nabla\phi)^2\Box\phi
  - 16\nabla_\mu\phi\nabla_\nu\phi\nabla^\mu\nabla^\nu\phi
  - 8\nabla_\mu\nabla_\nu\phi\,\nabla^\mu\nabla^\nu\phi = 0,
  \label{eq:scalar_eom}
\end{equation}
\end{widetext}
where $\Box \equiv \nabla^\mu\nabla_\mu$ denotes the covariant
d'Alembertian operator. Variation with respect to the metric $g_{\mu\nu}$
leads to the modified gravitational field equations
\begin{equation}
  E_{\mu\nu} = G_{\mu\nu} + \alpha\, H_{\mu\nu} = \kappa\, T_{\mu\nu},
  \label{eq:field_eqs}
\end{equation}
where $T_{\mu\nu}$ is the matter energy--momentum tensor and $H_{\mu\nu}$
collectively denotes the higher-curvature scalar--tensor contributions
arising from the Gauss--Bonnet coupling. The purely geometric
Gauss--Bonnet tensor entering $H_{\mu\nu}$ is given by
\begin{equation}
\begin{split}
  H_{\mu\nu} = 2\Bigl(&RR_{\mu\nu} - 2R_{\mu\alpha\nu\beta}R^{\alpha\beta}
  + R_{\mu\alpha\beta\sigma}R_{\nu}{}^{\alpha\beta\sigma}\\
  &- 2R_{\mu\alpha}R^{\alpha}{}_\nu
  - \tfrac{1}{4}g_{\mu\nu}\mathcal{G}\Bigr).
\end{split}
  \label{eq:GB_tensor}
\end{equation}
An important consistency relation between the scalar and metric field
equations is provided by the on-shell trace identity,
\begin{equation*}
  \kappa\, g^{\mu\nu}T_{\mu\nu} = -R - \frac{\alpha}{2}\mathcal{G},
\end{equation*}
which serves as a useful diagnostic for verifying solutions but is not
required for the stellar structure derivation that follows.

\subsection{Equilibrium Configurations via Modified TOV Equations}
\label{subsec:tov}

Our aim is to characterise static equilibrium configurations of compact
quark stars within the regularized 4DEGB gravitational framework.
Accordingly, we restrict the field equations to spacetimes possessing
static spherical symmetry. The internal geometry is parametrised by the
line element
\begin{equation}
  ds^2 = -e^{2\Phi(r)}dt^2 + e^{2\Lambda(r)}dr^2 + r^2\,d\Omega^2,
  \label{eq:metric}
\end{equation}
where the metric potentials $\Phi(r)$ and $\Lambda(r)$ depend solely
on the areal radius $r$.

The matter content of the stellar interior is described as a single
perfect fluid of deconfined quark matter. The energy--momentum tensor
takes the standard form
\begin{equation}
  T_{\mu\nu} = (\varepsilon + p)\,u_\mu u_\nu + p\,g_{\mu\nu},
  \label{eq:Tmunu}
\end{equation}
where $\varepsilon$ and $p$ denote the energy density and isotropic
pressure of the quark matter, respectively, and $u^\mu$ is the
four-velocity normalised by $u^\mu u_\mu = -1$. Throughout this work
the stellar interior is composed entirely of deconfined quark matter,
so $\varepsilon$ and $p$ in Eq.~\eqref{eq:Tmunu} are determined
directly by the FKV[$X$] equation of state described in
Sec.~\ref{sec:eos}.
Substituting the metric ansatz~\eqref{eq:metric} together with the
energy--momentum tensor~\eqref{eq:Tmunu} into the 4DEGB field
equations yields the $tt$ and $rr$ components in the form
\begin{widetext}
\begin{align}
  \frac{2}{r}\frac{d\Lambda}{dr} &= e^{2\Lambda}
  \left[8\pi\varepsilon -
  \frac{1-e^{-2\Lambda}}{r^2}
  \!\left(1 - \frac{\alpha(1-e^{-2\Lambda})}{r^2}\right)\right]
  \!\left(1 + \frac{2\alpha(1-e^{-2\Lambda})}{r^2}\right)^{\!-1},
  \label{eq:tt_comp}\\[4pt]
  \frac{2}{r}\frac{d\Phi}{dr} &= e^{2\Lambda}
  \left[8\pi p +
  \frac{1-e^{-2\Lambda}}{r^2}
  \!\left(1 - \frac{\alpha(1-e^{-2\Lambda})}{r^2}\right)\right]
  \!\left(1 + \frac{2\alpha(1-e^{-2\Lambda})}{r^2}\right)^{\!-1}.
  \label{eq:rr_comp}
\end{align}
\end{widetext}
The local conservation of energy--momentum, $\nabla_\mu T^{\mu\nu} = 0$,
yields the hydrostatic equilibrium condition
\begin{equation}
  \frac{dp}{dr} = -(\varepsilon + p)\frac{d\Phi}{dr}.
  \label{eq:hydrostatic}
\end{equation}
Introducing the gravitational mass function $m(r)$ via
\begin{equation}
  e^{-2\Lambda(r)} = 1 + \frac{r^2}{2\alpha}
  \!\left(1 - \sqrt{1 + \frac{8\alpha\,m(r)}{r^3}}\right),
  \label{eq:mass_func}
\end{equation}
one recovers the Schwarzschild limit as $\alpha \to 0$.
The mass function satisfies
\begin{equation}
  \frac{dm}{dr} = 4\pi r^2\,\varepsilon.
  \label{eq:dmdr}
\end{equation}
Combining Eqs.~\eqref{eq:rr_comp}, \eqref{eq:hydrostatic},
and~\eqref{eq:mass_func}, and defining
\begin{equation}
  \Gamma(r) \equiv \sqrt{1 + \frac{8\alpha\,m(r)}{r^3}},
  \label{eq:Gamma_def}
\end{equation}
we obtain the modified Tolman--Oppenheimer--Volkoff equation governing
hydrostatic equilibrium for a quark star in 4DEGB gravity:
\begin{equation}
  \frac{dp}{dr} =
  \frac{(\varepsilon + p)
        \bigl[r^3\bigl(\Gamma + 8\pi\alpha\,p - 1\bigr) - 2\alpha\,m\bigr]}
       {r^2\,\Gamma\bigl[r^2(\Gamma - 1) - 2\alpha\bigr]}.
  \label{eq:tov_full}
\end{equation}
Equations~\eqref{eq:dmdr} and~\eqref{eq:tov_full} together constitute
the complete stellar structure system for static equilibrium
configurations of quark stars in regularized 4DEGB gravity. In the
limit $\alpha \to 0$ one has $\Gamma \to 1$, and
Eq.~\eqref{eq:tov_full} reduces exactly to the standard
Tolman--Oppenheimer--Volkoff equation of general relativity.

Throughout this work we adopt natural units with $\hbar = G = c = 1$.
All quantities are expressed in geometrised units with lengths in
kilometres, so that masses carry the conversion
$M_\odot = 1.4766\,\mathrm{km}$.

\subsection{Buchdahl Bound, Black Hole Horizon, and Observational
            Constraints}
\label{subsec:constraints}

The vacuum solution to the field equations is given by
Eq.~\eqref{eq:mass_func} with $m(r) = M = \mathrm{const}$, describing
a black hole with outer horizon radius~\cite{Gammon:2023uss}
\begin{equation}
  R_h = M + \sqrt{M^2 - \alpha},
  \label{eq:horizon}
\end{equation}
provided $M \geq M_{\rm BH}^{\min} = \sqrt{\alpha}$. Unlike GR, the
4DEGB theory lacks a mass gap between compact stars and black holes of
the same radius; the mass--radius curves of stellar sequences
asymptotically approach $R_h(M)$ for sufficiently large
$\alpha$~\cite{Charmousis:2021npl,Gammon:2023uss}.

The Buchdahl bound has been derived in 4DEGB gravity
in~\cite{PhysRev.116.1027,Gammon:2023uss} using arguments analogous to
those in GR, yielding
\begin{equation}
  \sqrt{1 - \mu R^2}\,(1 + \alpha\mu) > \frac{1}{3}(1 - \alpha\mu),
  \label{eq:buchdahl}
\end{equation}
where
\begin{equation*}
  \mu \equiv \frac{1}{2\alpha}
  \!\left(\sqrt{1 + \frac{8M\alpha}{R^3}} - 1\right).
\end{equation*}
For small $\alpha$ this reduces to $M/R \leq 4/9 + (16/27)\,\alpha/R^2$,
recovering the GR Buchdahl limit $C \leq 4/9$ as $\alpha \to 0$.
For $\alpha > 0$ the bound is less restrictive and depends on both $M$
and $R$ along the stellar sequence; it is discussed in the text
alongside the compactness results.

Regarding observational constraints on the coupling, a study of LAGEOS
satellite data and cosmological
observations~\cite{Clifton:2020xhc,Fernandes:2022zrq} yields the broad
bound $\alpha \lesssim 10^{10}\,\mathrm{m}^2 = 10^4\,\mathrm{km}^2$.
Preliminary analyses of gravitational-wave and compact-object data suggest
a considerably tighter constraint $\alpha \lesssim 10^7\,\mathrm{m}^2 =
10\,\mathrm{km}^2$~\cite{Gammon:2023uss,Charmousis:2021npl}. In the
present work we adopt $\alpha \in \{0, 1, 10\}\,\mathrm{km}^2$ as our
parameter grid, motivated by this tighter GW-oriented bound. The value
$\alpha = 0$ recovers the GR baseline, $\alpha = 1\,\mathrm{km}^2$ lies
well within the constrained regime, and $\alpha = 10\,\mathrm{km}^2$ sits
at the boundary of the compact-object bound, thereby probing the maximally
allowed deviation from GR within the current observational constraints.

\begin{figure*}[ht]
  \centering
  \includegraphics[width=0.75\textwidth]{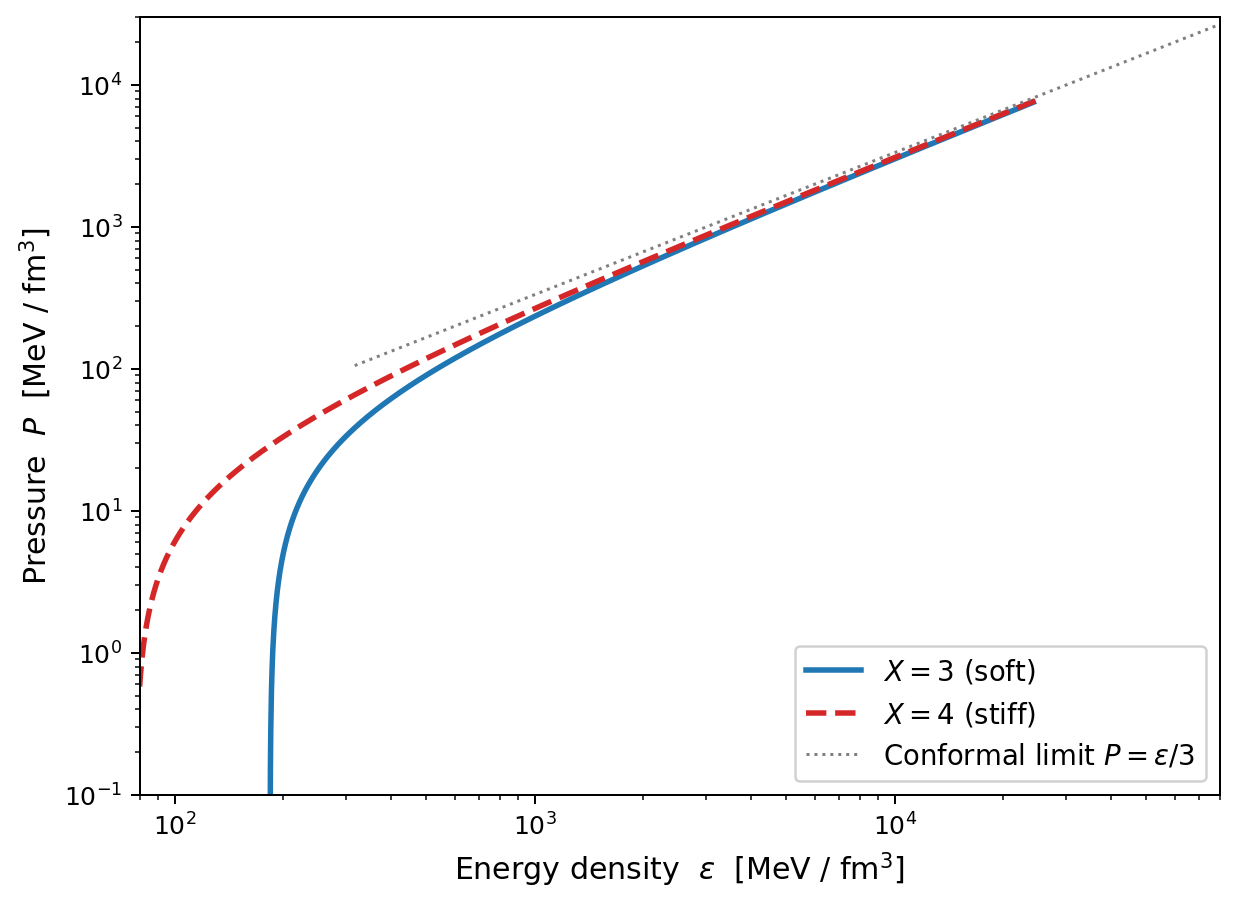}
  \caption{Pressure $P$ as a function of energy density $\varepsilon$
           for the FKV[$X$] equation of state, evaluated for
           $X = 3$ (soft, solid blue) and $X = 4$ (stiff, dashed red).
           The dotted gray line shows the conformal limit
           $P = \varepsilon/3$. Both curves originate at their
           respective surface energy densities where $P = 0$, and
           approach but remain below the conformal limit throughout
           the domain $\muB \leq 2.0\,\mathrm{GeV}$.
           Units are $\mathrm{MeV\,fm}^{-3}$ on both axes.}
  \label{fig:eos_P_eps}
\end{figure*}

\begin{figure*}[ht]
  \centering
  \includegraphics[width=0.75\textwidth]{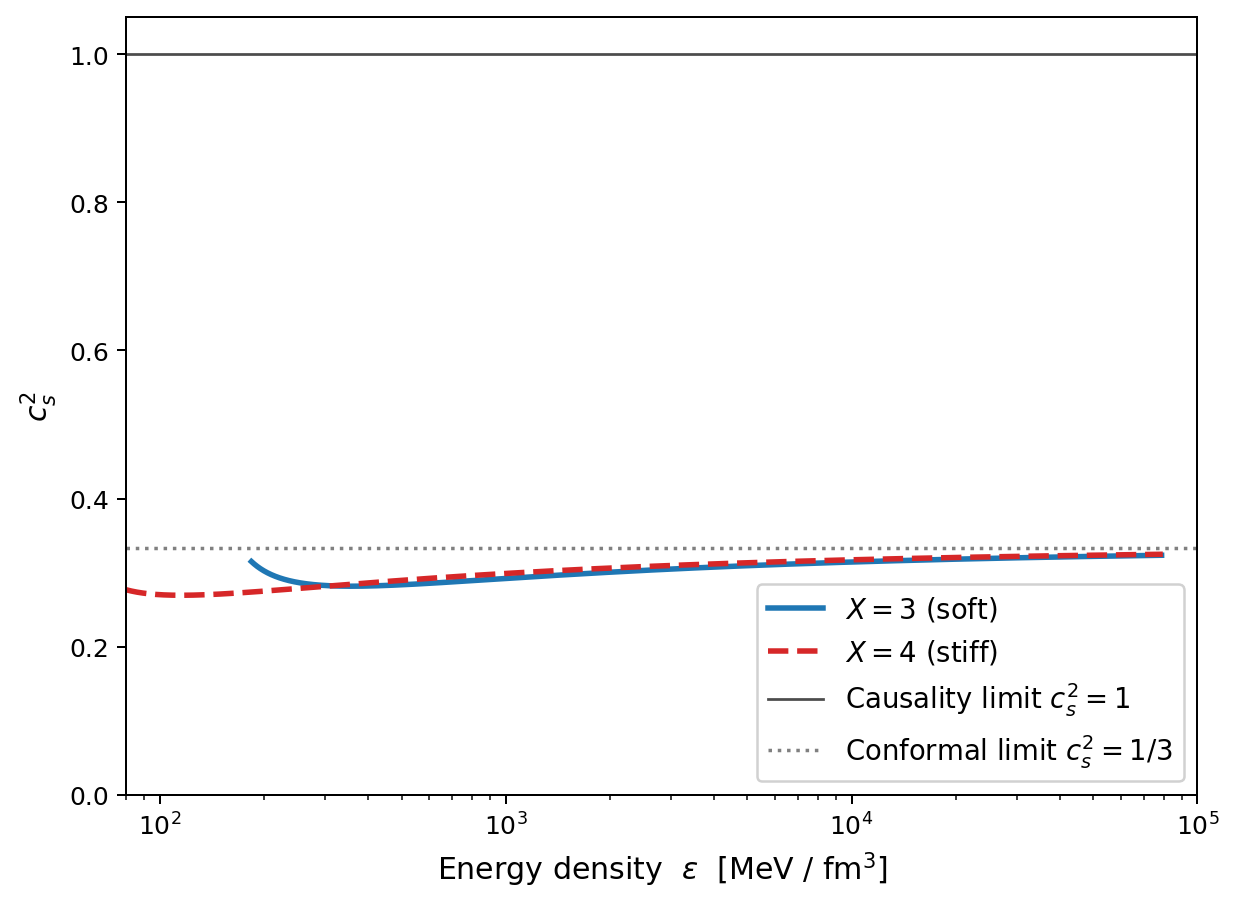}
  \caption{Squared speed of sound $\cs^2$ as a function of energy
           density $\varepsilon$ for the FKV[$X$] EOS with
           $X = 3$ (soft, solid blue) and $X = 4$ (stiff, dashed red).
           The solid gray horizontal line marks the causality limit
           $\cs^2 = 1$ and the dotted gray line the conformal limit
           $\cs^2 = 1/3$. Both curves remain sub-conformal and causal
           throughout the entire domain. The asymptotic approach
           to $\cs^2 \to 1/3$ as $\varepsilon \to \infty$ is consistent
           with the perturbative QCD prediction for massless quark matter.}
  \label{fig:eos_cs2}
\end{figure*}
\section{Cold Quark Matter from Perturbative QCD}
\label{sec:eos}

\subsection{Pocket Formula}
\label{subsec:pocket}

The pressure of cold, beta-equilibrated quark matter at baryon
chemical potential $\muB$ is obtained from the three-loop pQCD
result of Kurkela, Romatschke and Vuorinen~\cite{Kurkela:2010cold},
which Fraga, Kurkela and Vuorinen~\cite{Fraga:2014interacting} cast
in the simple analytic form
\begin{equation}
  P(\muB, X) = P_{\mathrm{SB}}(\muB)
  \!\left[c_1 - \frac{a(X)}{\muB/\mathrm{GeV} - b(X)}\right],
  \label{eq:fkv}
\end{equation}
where the Stefan--Boltzmann pressure of three massless flavours is
\begin{equation}
  P_{\mathrm{SB}}(\muB) = \frac{3}{4\pi^2}
  \!\left(\frac{\muB}{3}\right)^{\!4}
  = \frac{\muB^4}{108\pi^2},
  \label{eq:psb}
\end{equation}
Throughout this section all thermodynamic quantities are expressed in
natural units ($\hbar = c = 1$); the conversion to
$\mathrm{MeV\,fm}^{-3}$ used in Figs.~\ref{fig:eos_P_eps}
and~\ref{fig:eos_cs2} is
$P\,[\mathrm{MeV\,fm}^{-3}] = P\,[\mathrm{GeV}^4]\times
10^3/(\hbar c)^3$, where $\hbar c = 0.197327\,\mathrm{GeV\,fm}$.
The scale-dependent fit functions are
\begin{equation}
  a(X) = d_1\,X^{-\nu_1}, \qquad
  b(X) = d_2\,X^{-\nu_2}.
  \label{eq:ab}
\end{equation}
The best-fit constants are~\cite{Fraga:2014interacting}
\begin{equation}
\begin{split}
  &c_1 = 0.9008,\quad d_1 = 0.5034,\quad d_2 = 1.452,\\
  &\nu_1 = 0.3553,\quad \nu_2 = 0.9101.
\end{split}
  \label{eq:constants}
\end{equation}
The dimensionless parameter $X$ encodes the renormalization-scale
ambiguity of the perturbative QCD calculation, with $X \in [1,4]$
spanning the range over which the renormalization point is varied
relative to the quark chemical potential. We focus on $X = 3$ and
$X = 4$, two representative values in the upper part of the FKV
renormalization-scale range that together bracket the theoretical
uncertainty band: $X = 3$ yields a softer EOS with a higher surface
energy density, while $X = 4$ yields a stiffer EOS with a lower surface
energy density and correspondingly larger maximum masses. Lower values
$X \lesssim 2$ produce surface chemical potentials close to the pole
of the rational factor in Eq.~\eqref{eq:fkv}, leading to acausal or
numerically unreliable behavior, and are not considered here. The
pocket formula is valid for $\muB \lesssim 2\,\mathrm{GeV}$,
which defines the upper limit of the integration in the TOV
solver; results above this limit lie outside the domain of
applicability of the perturbative expansion~\cite{Fraga:2014interacting}.
It should be noted that the FKV[$X$] expression is used
phenomenologically over the full stellar profile, including the
low-density surface region where $\muB \approx \muB^{(0)}(X)$; this
region lies at the lower edge of the controlled perturbative domain,
and the associated uncertainty is subsumed into the renormalization-scale
band spanned by $X \in \{3,4\}$.

\subsection{Thermodynamic Quantities}
\label{subsec:thermo}

The baryon number density is obtained by differentiating
Eq.~\eqref{eq:fkv} with respect to $\muB$ at fixed $X$, using the
Gibbs-Duhem relation $n_B = \partial P / \partial \muB$:
\begin{equation}
  n_B(\muB, X) = \frac{4P(\muB, X)}{\muB}
  + \frac{P_{\mathrm{SB}}(\muB)\,a(X)}{(\muB/\mathrm{GeV} - b(X))^2},
  \label{eq:nB}
\end{equation}
where the second term arises from differentiating the rational factor
in Eq.~\eqref{eq:fkv}. The energy density follows from the zero-temperature
Euler relation,
\begin{equation}
  \varepsilon(\muB, X) = \muB\,n_B(\muB, X) - P(\muB, X).
  \label{eq:eps}
\end{equation}
The surface of the star is defined by the condition $P(\muB^{(0)}, X) = 0$,
which yields the surface chemical potential
\begin{equation}
  \muB^{(0)}(X) = \mathrm{GeV}\times\!\left[b(X) + \frac{a(X)}{c_1}\right].
  \label{eq:muB_surf}
\end{equation}
For $X = 3$ this gives $\muB^{(0)} \approx 0.913\,\mathrm{GeV}$, while
for $X = 4$ one finds $\muB^{(0)} \approx 0.753\,\mathrm{GeV}$.
Since $P = 0$ at $\muB^{(0)}$ and the energy density is finite and
positive there, the FKV[$X$] EOS describes a selfbound fluid: quark
matter is stable against decompression without the need for an external
bag pressure parameter. The pressure-energy density relation is shown in
Fig.~\ref{fig:eos_P_eps} for both values of $X$.

\subsection{Speed of Sound and Causality}
\label{subsec:cs2}

The adiabatic speed of sound squared is obtained via parametric
differentiation with respect to $\muB$,
\begin{equation}
  \cs^2(\muB, X) = \frac{dP}{d\varepsilon}
  = \frac{dP/d\muB}{d\varepsilon/d\muB}
  = \frac{n_B(\muB, X)}{\muB\,dn_B/d\muB},
  \label{eq:cs2}
\end{equation}
where the last equality follows from $dP/d\muB = n_B$ and
$d\varepsilon/d\muB = \muB\,dn_B/d\muB$. The derivative
$dn_B/d\muB$ is computed analytically from Eq.~\eqref{eq:nB}:
\begin{equation}
  \frac{dn_B}{d\muB} = \frac{1}{108\pi^2}
  \!\left[12\muB^2 f + 8\muB^3 f' + \muB^4 f''\right],
  \label{eq:dnBdmu}
\end{equation}
where $f \equiv c_1 - a(X)/(\muB/\mathrm{GeV} - b(X))$,
$f' \equiv a(X)/(\muB/\mathrm{GeV} - b(X))^2$, and
$f'' \equiv -2a(X)/(\muB/\mathrm{GeV} - b(X))^3$.

The asymptotic behaviour of Eq.~\eqref{eq:cs2} is readily verified:
as $\muB \to \infty$ the perturbative corrections vanish and $f \to c_1$,
so that $P \to c_1 P_{\mathrm{SB}}$ and $\cs^2 \to 1/3$, recovering the
conformal limit expected for massless non-interacting quark matter.
Figure~\ref{fig:eos_cs2} displays $\cs^2$ as a function of energy
density for both values of $X$ over the full FKV validity domain. In
all cases the speed of sound satisfies $\cs^2 < 1/3$ throughout,
confirming that the FKV[$X$] EOS is everywhere sub-conformal. The
causality condition $\cs^2 < 1$ is satisfied with a substantial margin.
The non-monotonic feature near the surface in the $X = 3$ curve
reflects the proximity of the surface chemical potential to the
pole of the rational factor in Eq.~\eqref{eq:fkv}, and is a genuine
feature of the EOS rather than a numerical artefact.

\section{Numerical Results}
\label{sec:results}

%
%

We solve Eqs.~\eqref{eq:dmdr} and~\eqref{eq:tov_full} numerically for
$X \in \{3, 4\}$ and $\alphaGB \in \{0, 1, 10\}\,\mathrm{km}^2$, scanning
300 log-spaced central density values per $(X, \alphaGB)$ cell. For
each central density $\rhoC$, the stellar structure equations are
integrated radially outward from $r \to 0$ using a fourth-order
Runge--Kutta scheme, subject to the regularity condition $m(0) = 0$
and a prescribed central pressure. Integration is terminated when the
pressure vanishes, $p(R) = 0$, defining the stellar surface and the
total gravitational mass $M = m(R)$. The GR baseline ($\alphaGB = 0$)
reproduces the expected results to within $0.1\%$, confirming the
numerical consistency of the implementation.

\begin{figure*}[ht]
  \centering
  \begin{minipage}[t]{0.48\textwidth}
    \centering
    \includegraphics[width=\linewidth]{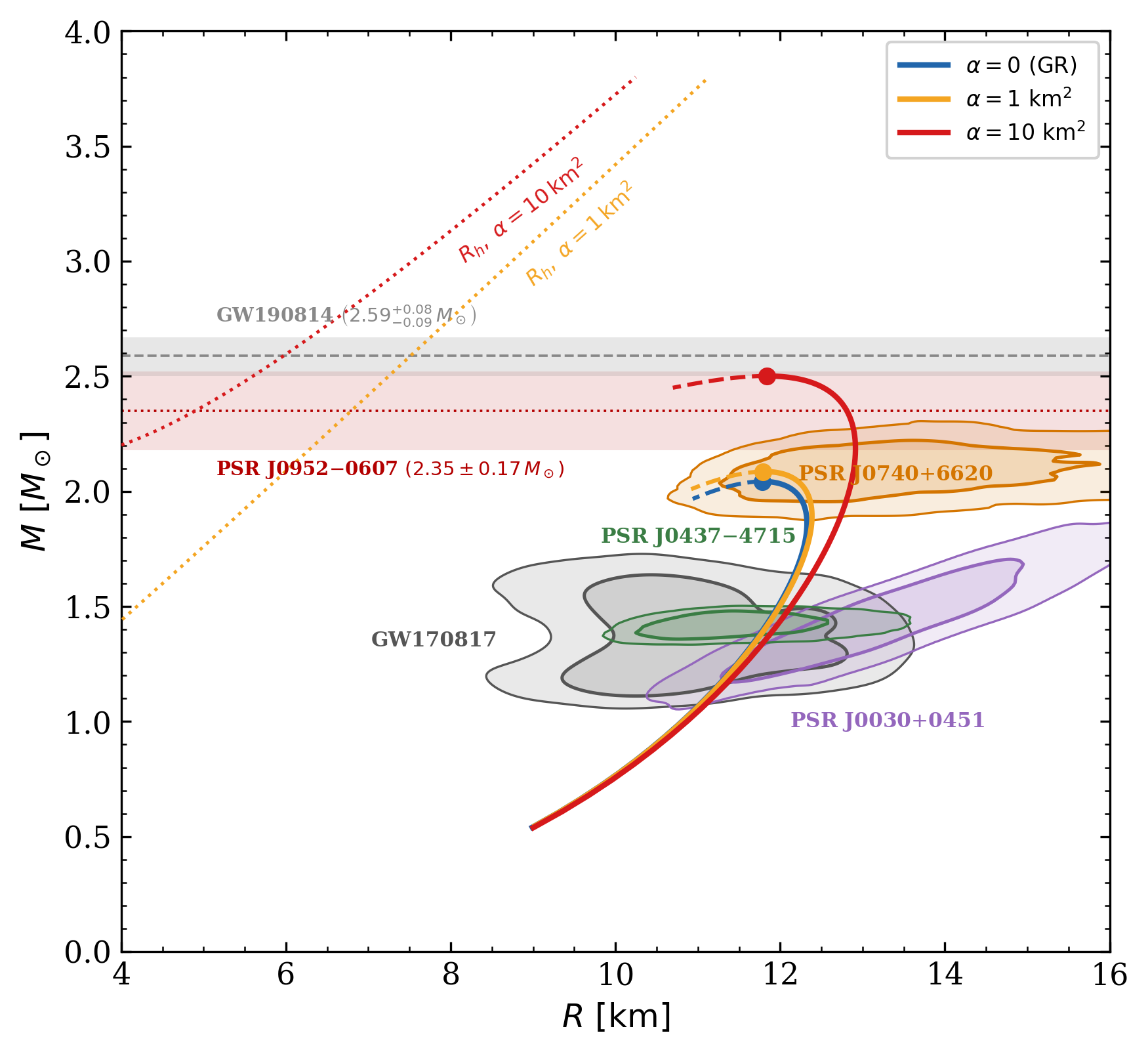}\\
    (a)
  \end{minipage}
  \hfill
  \begin{minipage}[t]{0.48\textwidth}
    \centering
    \includegraphics[width=\linewidth]{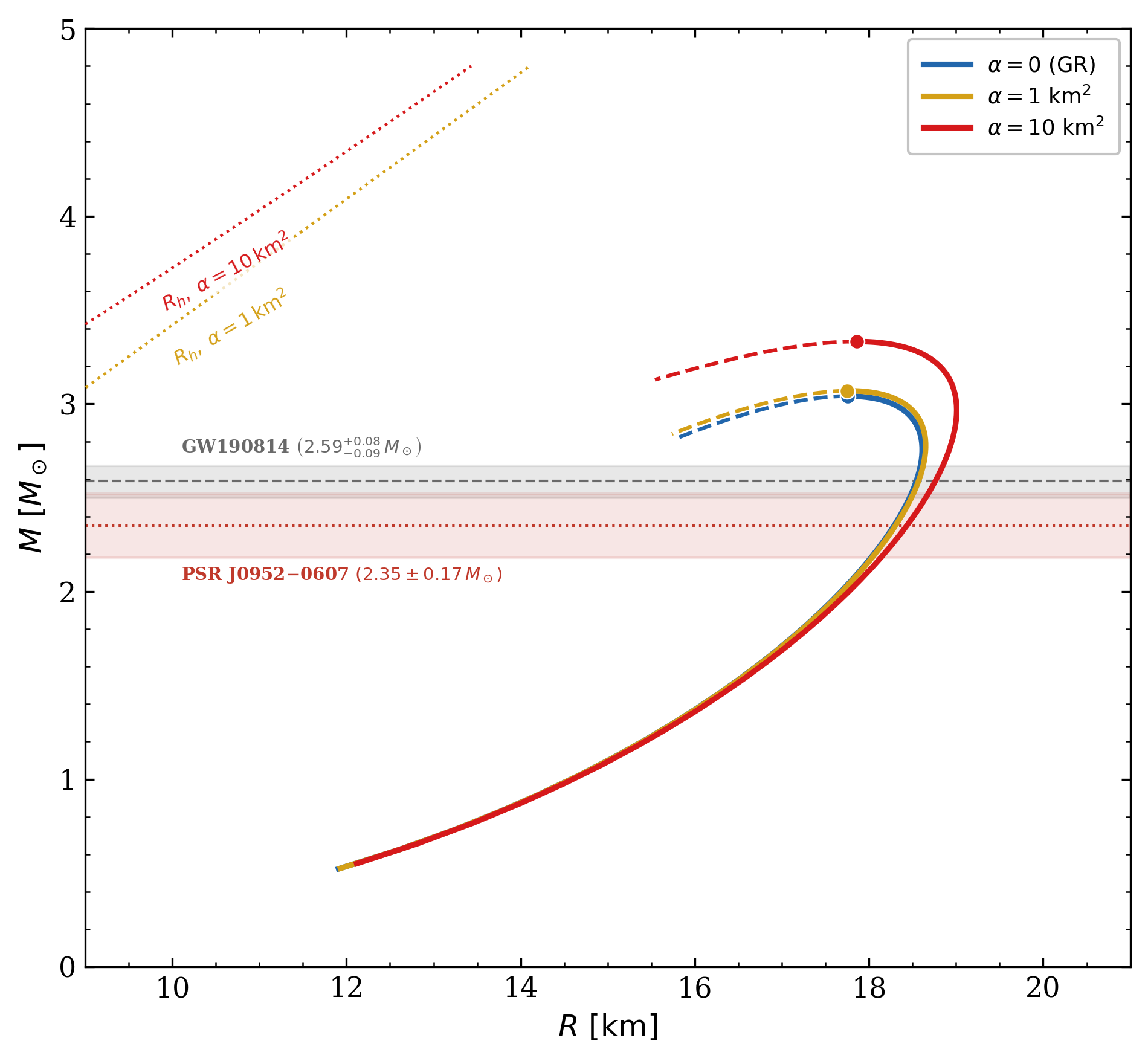}\\
    (b)
  \end{minipage}
  \caption{Mass--radius relations for quark stars in 4DEGB gravity
           with the FKV[$X$] EOS for $\alphaGB = 0$ (GR, blue),
           $1\,\mathrm{km}^2$ (gold), and $10\,\mathrm{km}^2$ (red):
           (a) $X = 3$ (soft); (b) $X = 4$ (stiff). Solid (dashed)
           curves indicate the stable (unstable) branch; filled circles
           denote the maximum-mass configuration. Dotted diagonal lines
           show the 4DEGB black hole horizon $R_h(\alphaGB)$ for
           $\alphaGB = 1$ and $10\,\mathrm{km}^2$. In panel (a), the
           gray dashed horizontal band marks the GW190814 secondary
           component
           ($2.59^{+0.08}_{-0.09}\,\Msun$~\cite{Abbott:2020gw190814})
           and the pink band the PSR~J0952$-$0607 mass
           ($2.35\pm0.17\,\Msun$~\cite{Romani:2022psr}); shaded
           contours show NICER constraints for
           PSR~J0740+6620~\cite{Fonseca:2021refined},
           PSR~J0437$-$4715~\cite{Choudhury:2024nicer}, and
           PSR~J0030+0451~\cite{Miller:2019psr}, together with the
           tidal constraint from GW170817~\cite{Abbott:2018gw170817}.
           These constraints are omitted from panel (b) for clarity
           (see text). In panel (b) all three configurations
           comfortably exceed the GW190814 and PSR~J0952$-$0607 mass
           constraints, with the stable branches nearly
           indistinguishable except near the maximum-mass turning
           point.}
  \label{fig:MR}
\end{figure*}

\subsection{Mass--Radius Relations}
\label{subsec:mr}

The mass--radius sequences for $X = 3$ and $X = 4$ are displayed in
panels (a) and (b) of Fig.~\ref{fig:MR}, respectively. In each
panel, solid curves denote the stable branch ($dM/d\rhoC > 0$) and
dashed curves the unstable branch ($dM/d\rhoC < 0$), with filled
circles marking the maximum-mass configuration for each value of
$\alphaGB$. The dotted diagonal lines in the upper-left region of each
panel indicate the 4DEGB black hole horizon radius
$R_h = M + \sqrt{M^2 - \alphaGB}$ for $\alphaGB = 1$ and
$10\,\mathrm{km}^2$; stellar configurations to the right of these lines
are compact objects distinct from black holes, while those lying on the
horizon line would collapse into a 4DEGB black hole.

For $X = 3$ [Fig.~\ref{fig:MR}(a)], the GR baseline
($\alphaGB = 0$) yields a maximum mass of
$\Mmax = 2.0431\,\Msun$ at radius $R = 11.78\,\mathrm{km}$.
Increasing the Gauss--Bonnet coupling to $\alphaGB = 1\,\mathrm{km}^2$
raises the maximum mass modestly to $2.0861\,\Msun$, while
$\alphaGB = 10\,\mathrm{km}^2$ produces a substantially larger
maximum mass of $2.5013\,\Msun$. This systematic enhancement reflects
the weakening of effective gravitational attraction induced by positive
higher-curvature corrections, which permits the stellar fluid to
support greater masses before reaching the onset of dynamical
instability. The radius at maximum mass increases only weakly with
$\alphaGB$, from $11.78$ to $11.83\,\mathrm{km}$, so that the primary
effect of the Gauss--Bonnet coupling is an upward shift of the M--R
curve rather than a lateral one.

Comparing with current observational constraints, the GR and
$\alphaGB = 1\,\mathrm{km}^2$ configurations fall below both the
PSR~J0952$-$0607 band ($2.35 \pm 0.17\,\Msun$~\cite{Romani:2022psr})
and the GW190814 secondary
($2.59^{+0.08}_{-0.09}\,\Msun$~\cite{Abbott:2020gw190814}). The
$\alphaGB = 10\,\mathrm{km}^2$ configuration, with
$\Mmax = 2.5013\,\Msun$, enters the lower portion of the
PSR~J0952$-$0607 band and approaches but does not reach the GW190814
constraint. The three M--R curves are consistent with the NICER
mass--radius contours for PSR~J0740+6620~\cite{Fonseca:2021refined,
Miller:2021radius,Riley:2021nicer},
PSR~J0437$-$4715~\cite{Choudhury:2024nicer}, and
PSR~J0030+0451~\cite{Miller:2019psr,Riley:2019nicer}, and with the
tidal deformability constraint from GW170817~\cite{Abbott:2018gw170817}.

For $X = 4$ [Fig.~\ref{fig:MR}(b)], the stiffer EOS produces
substantially larger maximum masses and radii across the board. The GR
baseline yields $\Mmax = 3.0415\,\Msun$ at $R = 17.75\,\mathrm{km}$,
already well above both the GW190814 and PSR~J0952$-$0607 constraints.
Increasing $\alphaGB$ to $1$ and $10\,\mathrm{km}^2$ raises $\Mmax$
further to $3.0698$ and $3.3324\,\Msun$, respectively. A notable
feature of the $X = 4$ sequences is that the three M--R curves are
nearly indistinguishable along the stable branch, separating only in
the immediate vicinity of $M_\text{max}$. This reflects the fact that
for the stiffer EOS the stellar structure is dominated by the lower
central densities characteristic of $X = 4$ stars, where the
Gauss--Bonnet correction $\Gamma - 1 \propto \alphaGB m / r^3$ is
comparatively small. The maximum radius also increases with $\alphaGB$,
from $17.75$ to $17.85\,\mathrm{km}$. The large stellar radii of $X = 4$ configurations place them well
outside the NICER contours shown in Fig.~\ref{fig:MR}(a), and we do
not overlay those constraints in panel~(b) to preserve clarity. We
note that radii in the range $17.7$--$17.9\,\mathrm{km}$ are
substantially larger than the canonical NICER constraints for
$\sim1.4\,\Msun$ neutron stars, which favour $R \lesssim
13\,\mathrm{km}$~\cite{Riley:2019nicer,Miller:2019psr}; this tension
would disfavour $X = 4$ as a description of typical compact objects
unless the FKV[4] EOS is interpreted as applying specifically to
selfbound strange quark stars rather than neutron stars.
\begin{figure*}[ht]
  \centering
  \begin{minipage}[t]{0.48\textwidth}
    \centering
    \includegraphics[width=\linewidth]{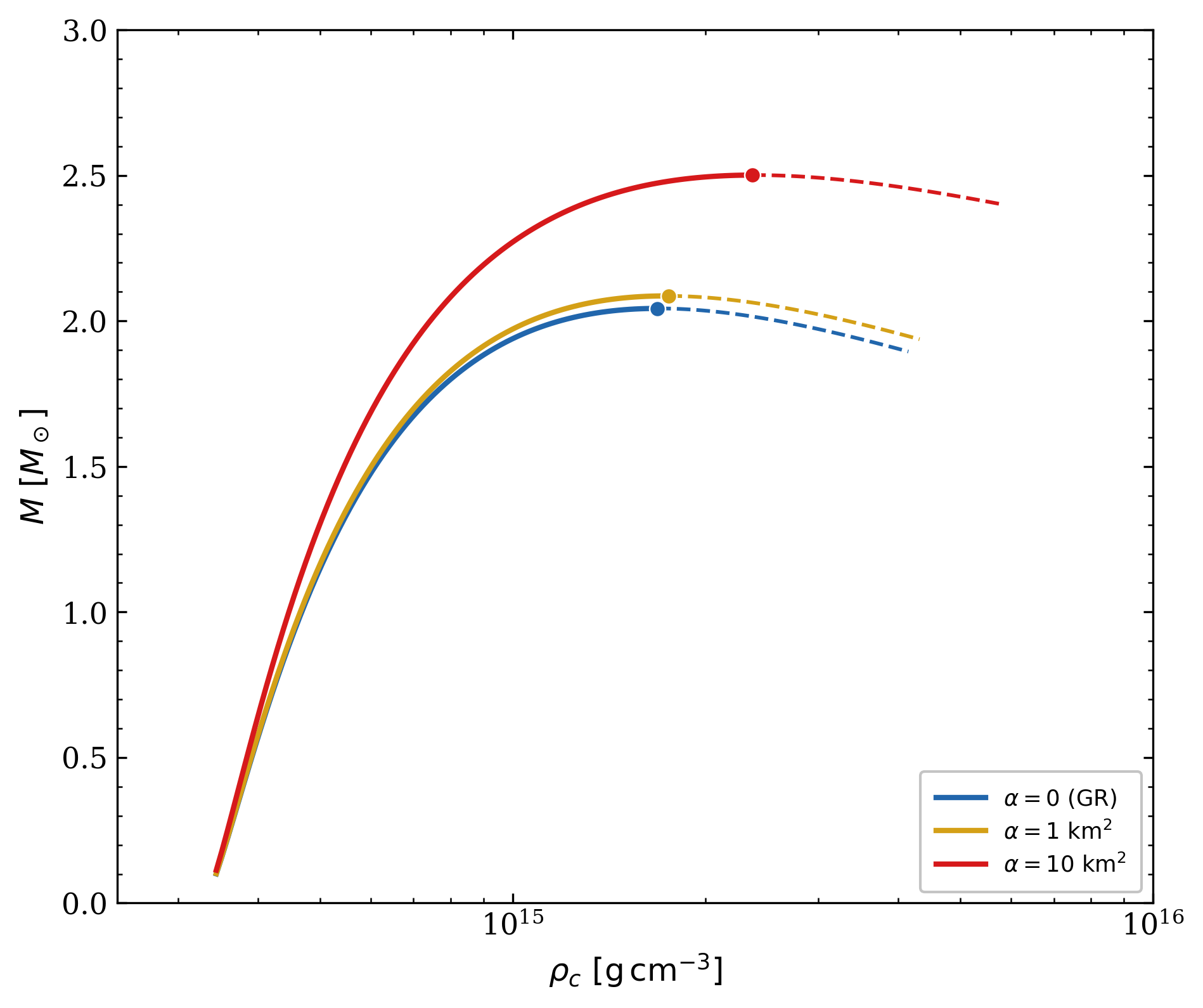}\\
    (a)
  \end{minipage}
  \hfill
  \begin{minipage}[t]{0.48\textwidth}
    \centering
    \includegraphics[width=\linewidth]{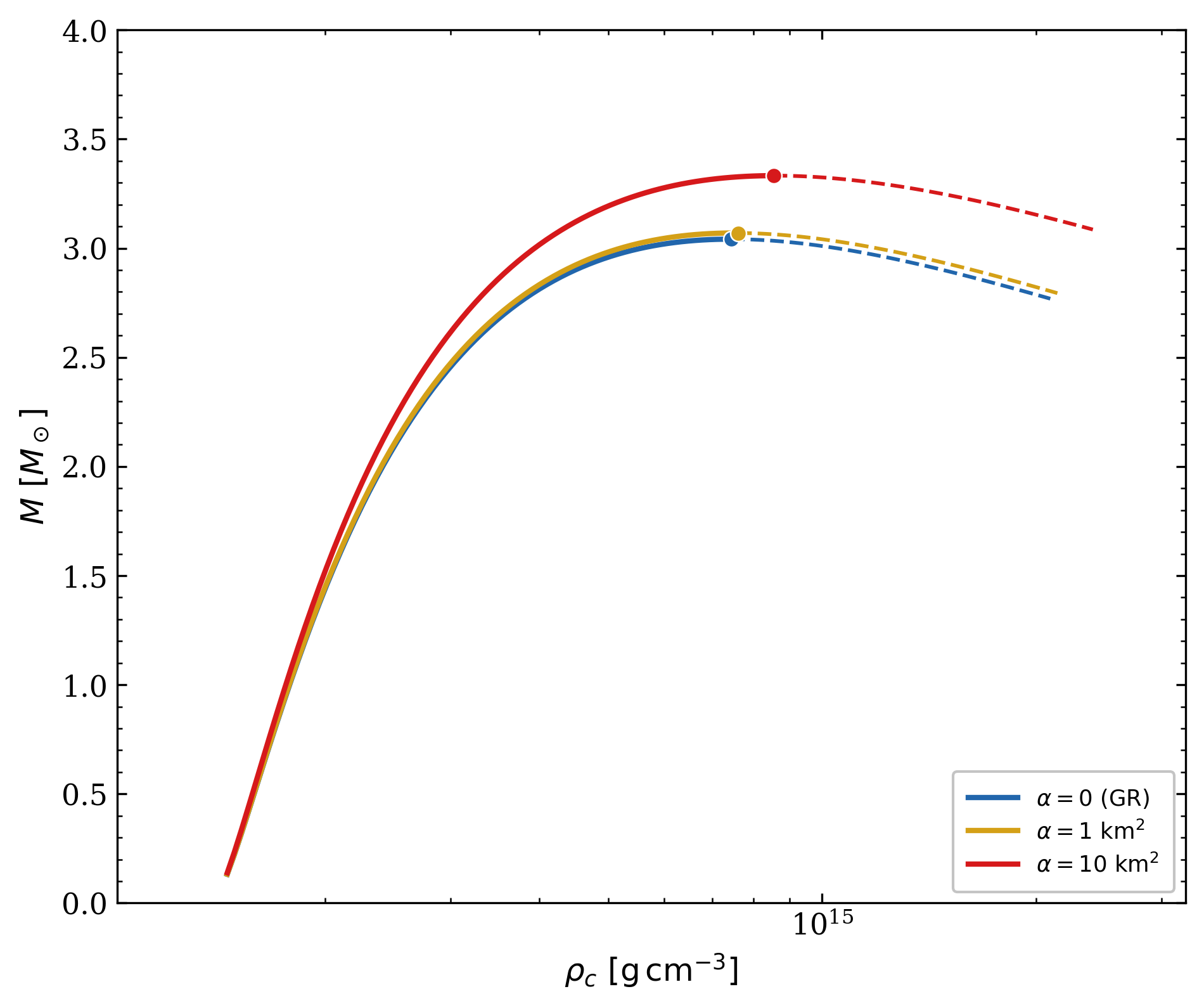}\\
    (b)
  \end{minipage}
  \caption{Gravitational mass $M$ versus central energy density $\rhoC$
           for $\alphaGB = 0$ (GR, blue), $1\,\mathrm{km}^2$ (gold),
           and $10\,\mathrm{km}^2$ (red): (a) $X = 3$; (b) $X = 4$.
           Solid (dashed) curves denote the stable (unstable) branch;
           filled circles mark the maximum-mass turning point.
           Increasing $\alphaGB$ shifts the turning point to higher
           $\rhoC$; for $X = 4$ the turning points lie at
           $\rhoC \sim 7$--$9\times10^{14}\,\mathrm{g\,cm}^{-3}$,
           significantly lower than for $X = 3$.}
  \label{fig:Mrho_pair}
\end{figure*}

\begin{figure*}[ht]
  \centering
  \begin{minipage}[t]{0.48\textwidth}
    \centering
    \includegraphics[width=\linewidth]{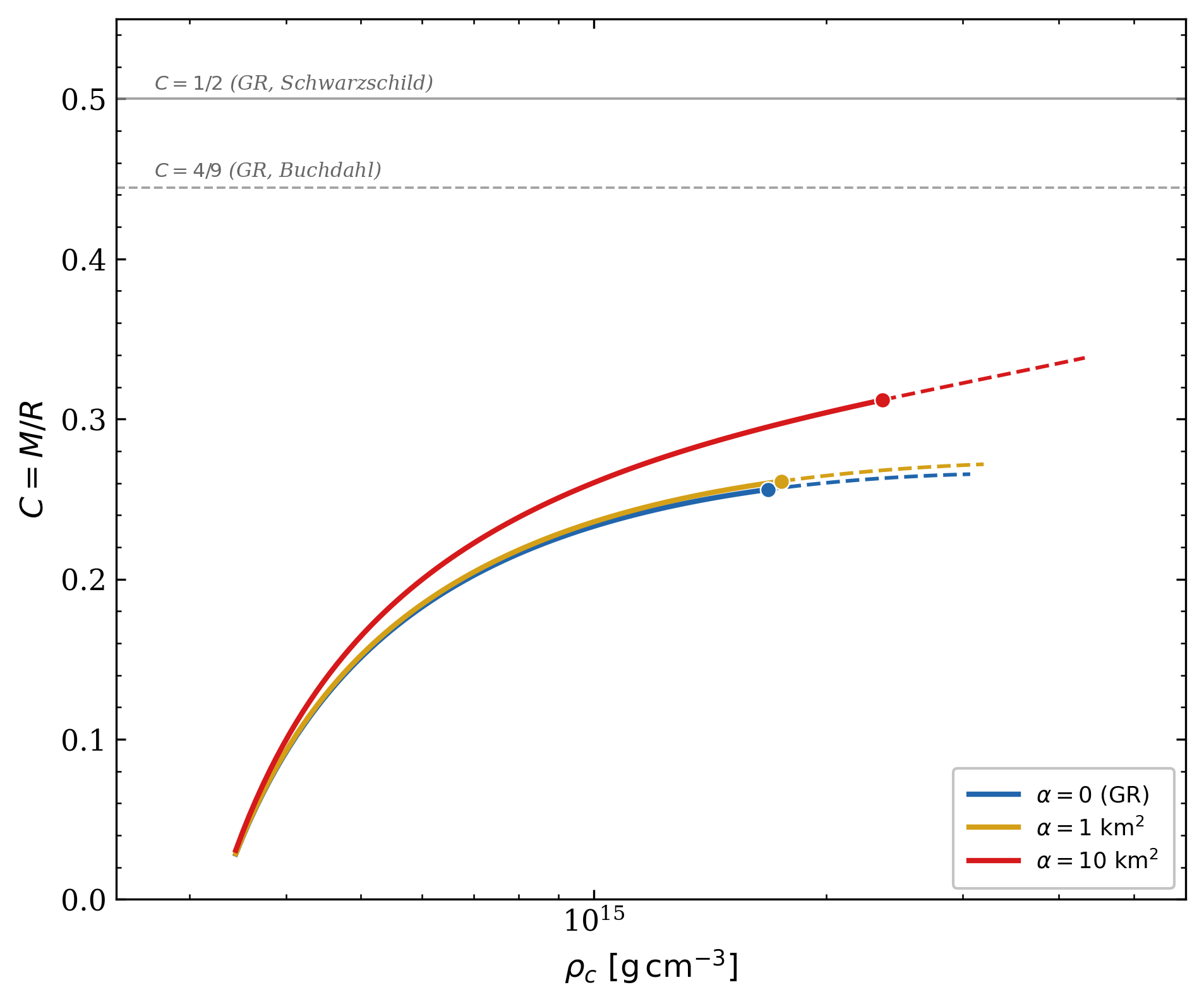}\\
    (a)
  \end{minipage}
  \hfill
  \begin{minipage}[t]{0.48\textwidth}
    \centering
    \includegraphics[width=\linewidth]{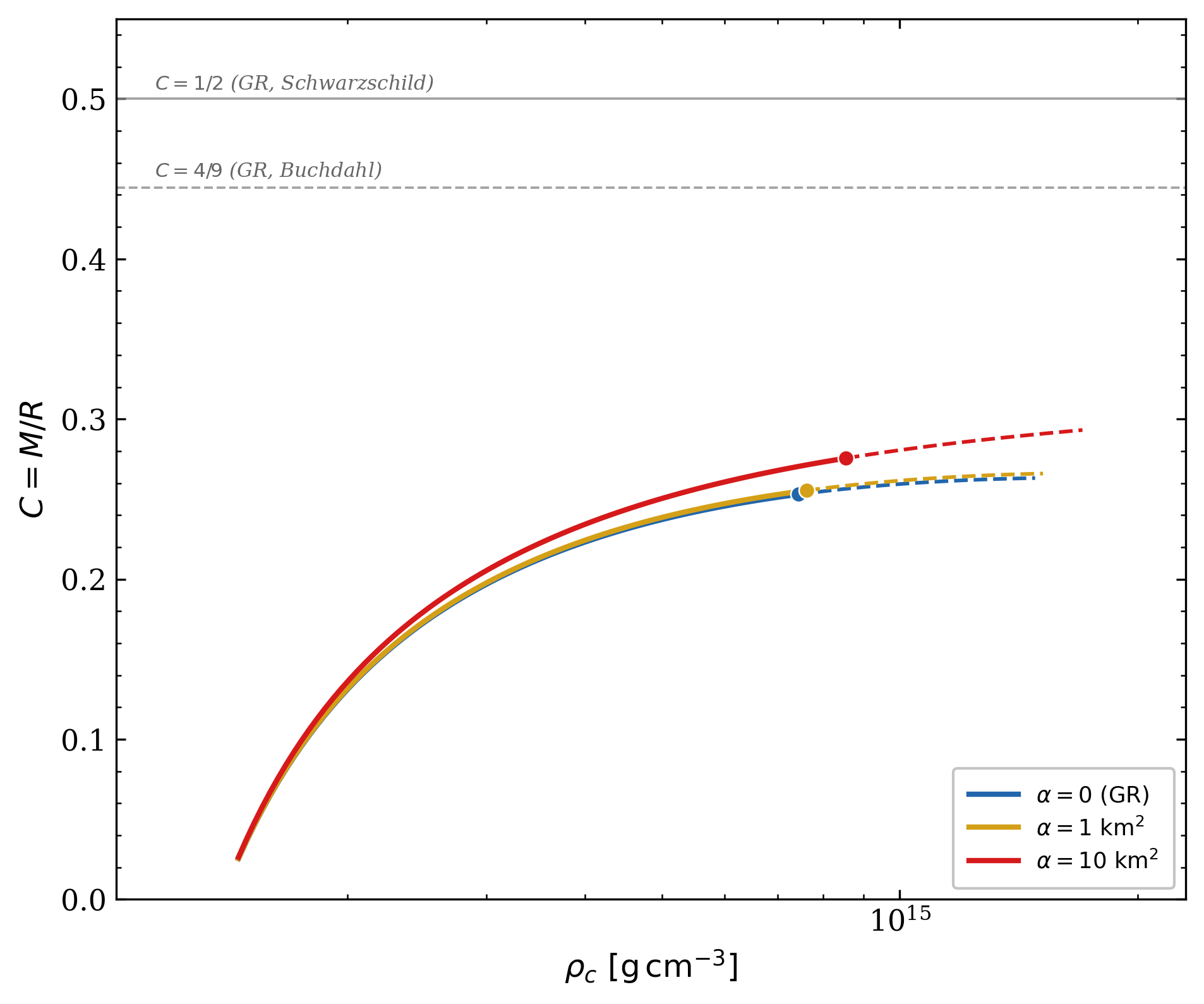}\\
    (b)
  \end{minipage}
  \caption{Compactness $C = M/R$ as a function of $\rhoC$ for
           $\alphaGB = 0$ (GR, blue), $1\,\mathrm{km}^2$ (gold), and
           $10\,\mathrm{km}^2$ (red): (a) $X = 3$; (b) $X = 4$.
           Solid (dashed) curves denote the stable (unstable) branch;
           filled circles mark $C$ at the maximum-mass configuration.
           The GR Schwarzschild limit ($C = 1/2$, solid gray) and
           Buchdahl bound ($C = 4/9$, dashed gray) are shown for
           reference. Note that the maximum of $C$ along each sequence
           does not coincide with $C(M_{\max})$ (see text). The $X = 4$
           configurations exhibit systematically lower compactness
           owing to their larger stellar radii; all configurations
           remain well below $C = 4/9$.}
  \label{fig:compact_pair}
\end{figure*}

\subsection{Mass--Central Density Sequences}
\label{subsec:mrho}

The gravitational mass as a function of central energy density
$\rhoC$ is shown in panels (a) and (b) of Fig.~\ref{fig:Mrho_pair}
for $X = 3$ and $X = 4$, respectively. These sequences provide the standard turning-point criterion for
radial stability: configurations on the rising branch
($dM/d\rhoC > 0$, solid curves) are identified as stable according
to this criterion, while those on the descending branch
($dM/d\rhoC < 0$, dashed curves) are interpreted as
unstable~\cite{Zeldovich:1971reas,Harrison:1965gtgc}. Filled circles
identify the maximum-mass turning point separating the two branches
for each value of $\alphaGB$.

For $X = 3$ [Fig.~\ref{fig:Mrho_pair}(a)], the GR and
$\alphaGB = 1\,\mathrm{km}^2$ sequences are nearly coincident on
the stable branch and reach their respective turning points at
central densities of $\rhoC \approx 1.68\times10^{15}$ and
$1.73\times10^{15}\,\mathrm{g\,cm}^{-3}$. The
$\alphaGB = 10\,\mathrm{km}^2$ sequence rises more steeply, reflecting
the enhanced mass support from the higher-curvature correction, and
reaches its maximum at a substantially higher central density of
$\rhoC \approx 2.38\times10^{15}\,\mathrm{g\,cm}^{-3}$.
The shift of the turning point to higher $\rhoC$ with increasing
$\alphaGB$ indicates that positive Gauss--Bonnet coupling shifts the
onset of dynamical instability to higher central densities, requiring
greater compression before the turning point is reached.

For $X = 4$ [Fig.~\ref{fig:Mrho_pair}(b)], the stiffer EOS produces
lower central densities throughout: the turning-point densities are
$\rhoC \approx 7.44\times10^{14}$, $7.54\times10^{14}$, and
$8.57\times10^{14}\,\mathrm{g\,cm}^{-3}$ for $\alphaGB = 0$, $1$, and
$10\,\mathrm{km}^2$, respectively. As with $X = 3$, the $\alpha = 0$
and $\alphaGB = 1\,\mathrm{km}^2$ sequences are indistinguishable on the
rising branch and only separate near the turning point, while
$\alphaGB = 10\,\mathrm{km}^2$ produces a visibly higher and more
extended stable branch. The complete set of turning-point parameters
is tabulated in Table~\ref{tab:results}.

\begin{figure*}[ht]
  \centering
  \begin{minipage}[t]{0.48\textwidth}
    \centering
    \includegraphics[width=\linewidth]{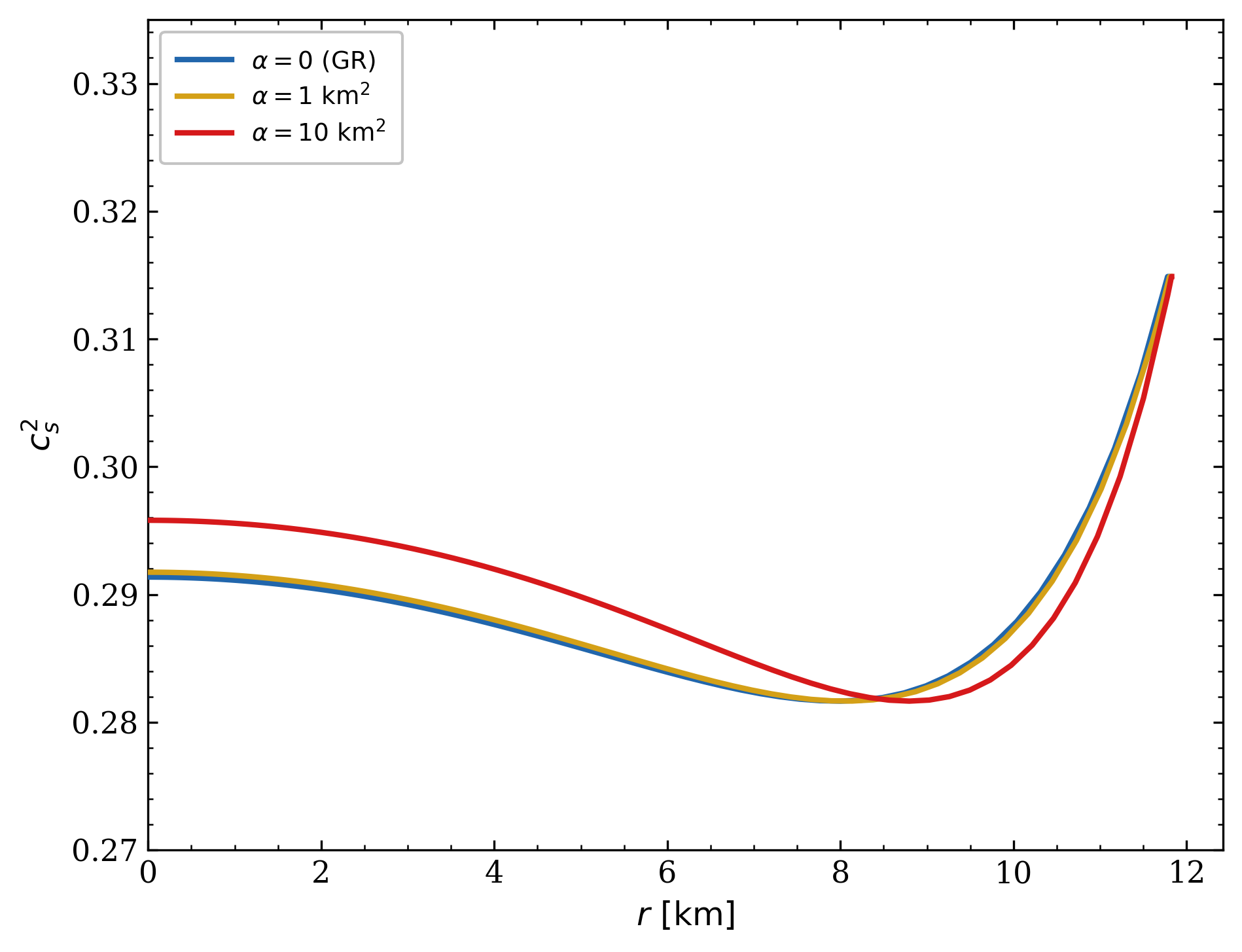}\\
    (a)
  \end{minipage}
  \hfill
  \begin{minipage}[t]{0.48\textwidth}
    \centering
    \includegraphics[width=\linewidth]{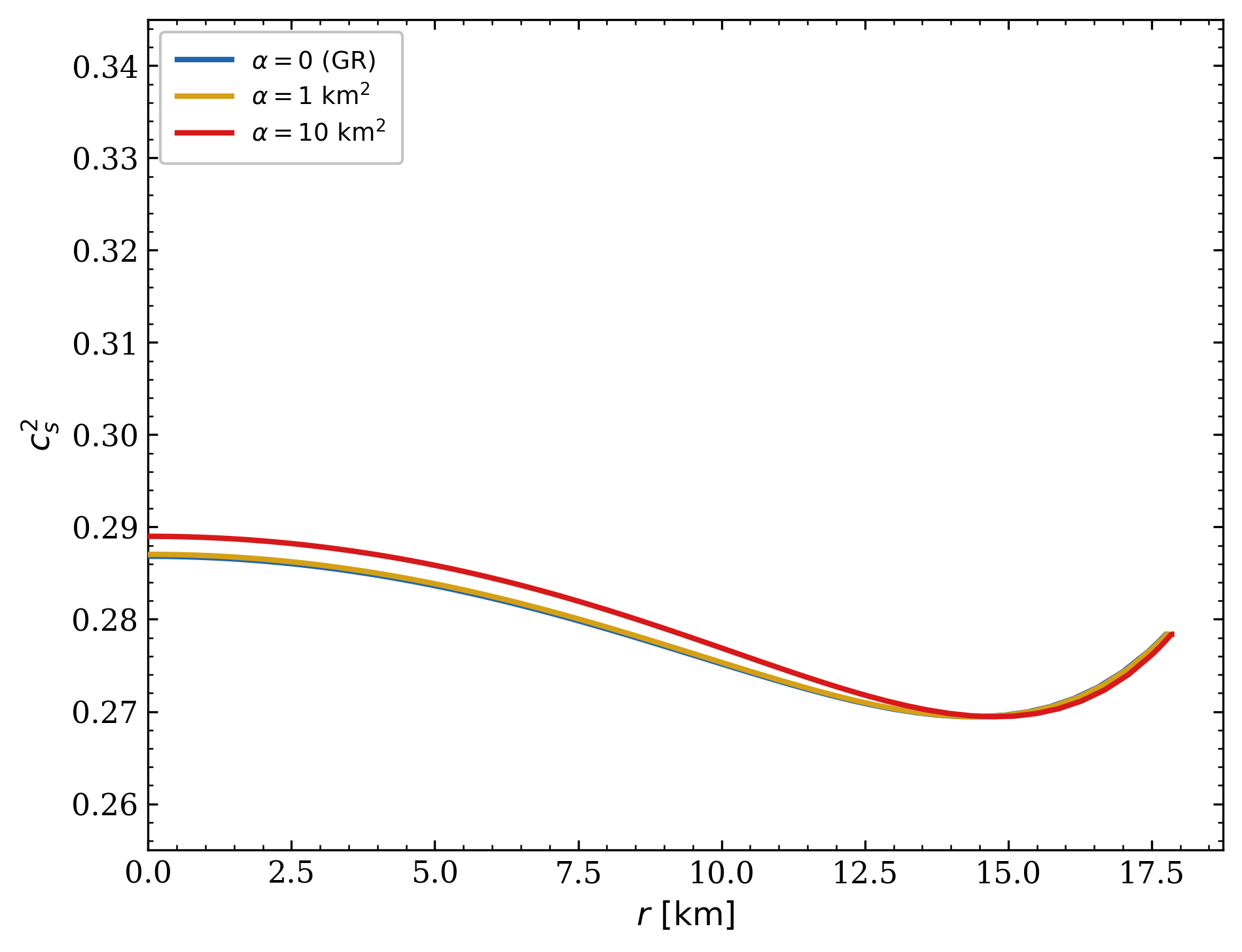}\\
    (b)
  \end{minipage}
  \caption{Radial profiles of the squared sound speed $\cs^2(r)$
           through the interior of the maximum-mass configuration for
           $\alphaGB = 0$ (GR, blue), $1\,\mathrm{km}^2$ (gold), and
           $10\,\mathrm{km}^2$ (red): (a) $X = 3$; (b) $X = 4$.
           The EOS itself is independent of $\alphaGB$; the separation
           between curves reflects the different
           $r \leftrightarrow \muB(r)$ mappings induced by the modified
           stellar structure. For $X = 3$, all profiles satisfy
           $\cs^2 < 1/3$ and converge to $\cs^2 \approx 0.315$ at the
           surface. For $X = 4$, the curves are nearly
           indistinguishable, with a minimum near
           $r \approx 14\,\mathrm{km}$ and a gentler surface upturn
           than for $X = 3$.}
  \label{fig:cs2_r_pair}
\end{figure*}

\begin{figure*}[ht]
  \centering
  \includegraphics[width=0.65\textwidth]{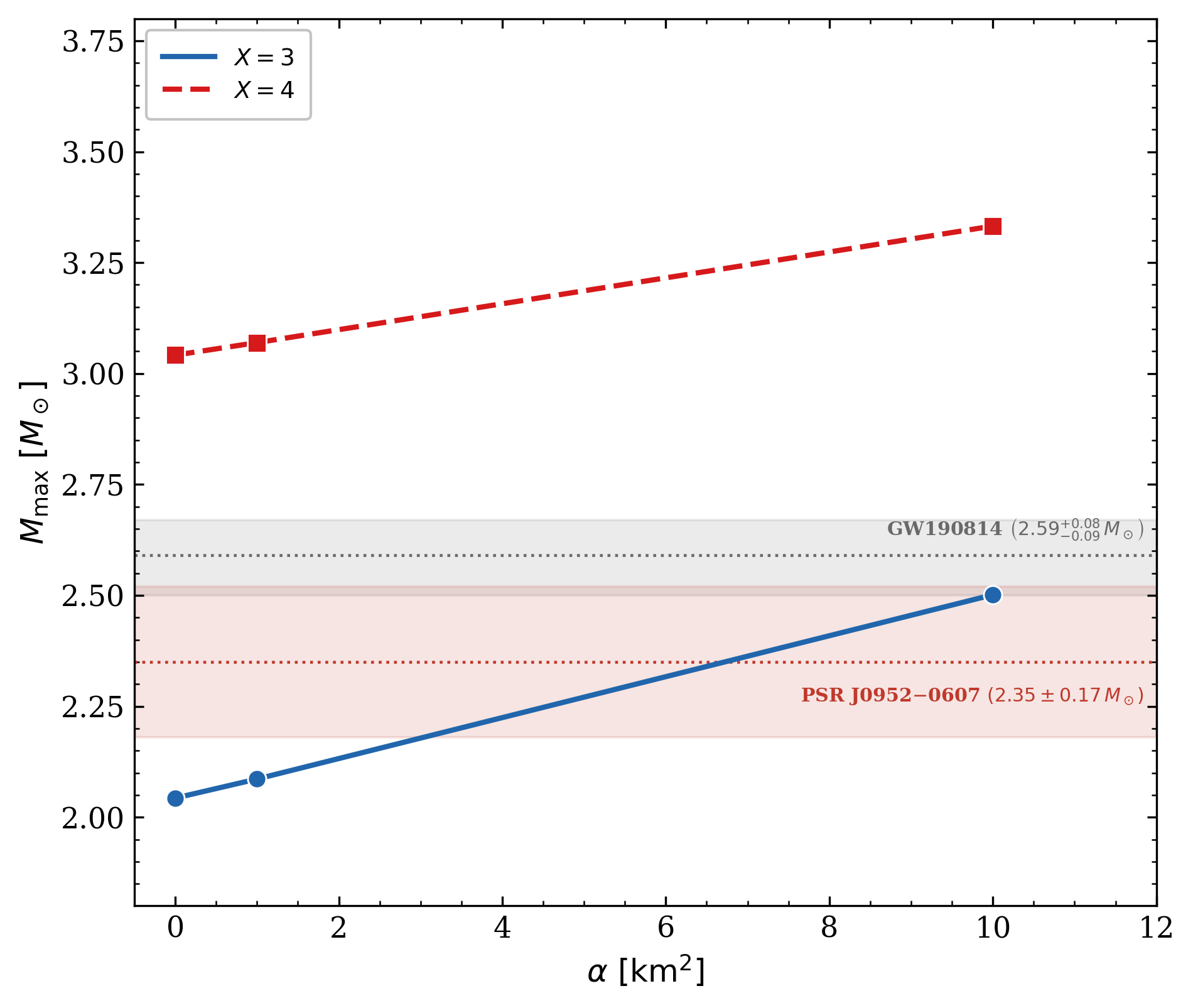}
  \caption{Maximum gravitational mass $\Mmax$ as a function of the
           Gauss--Bonnet coupling $\alphaGB$ for $X = 3$ (blue solid,
           circles) and $X = 4$ (red dashed, squares). Both sequences
           increase monotonically with $\alphaGB$. The gray shaded band
           indicates the GW190814 secondary component
           ($2.59^{+0.08}_{-0.09}\,\Msun$~\cite{Abbott:2020gw190814})
           and the pink band the PSR~J0952$-$0607 mass
           ($2.35 \pm 0.17\,\Msun$~\cite{Romani:2022psr}).
           For $X = 3$, only $\alphaGB = 10\,\mathrm{km}^2$ enters
           the PSR~J0952$-$0607 band. For $X = 4$, all configurations
           lie above both constraints.}
  \label{fig:Mmax_summary}
\end{figure*}

\subsection{Compactness}
\label{subsec:compact}

The compactness $C = M/R$ as a function of central density $\rhoC$
is displayed in panels (a) and (b) of Fig.~\ref{fig:compact_pair}
for $X = 3$ and $X = 4$, respectively. Two GR reference lines are
shown in each panel: the Schwarzschild limit $C = 1/2$ (solid gray)
and the Buchdahl bound $C = 4/9 \approx 0.444$~\cite{PhysRev.116.1027}
(dashed gray), which is the maximum compactness permitted for a perfect
fluid sphere in GR. As discussed in Sec.~\ref{subsec:constraints},
the 4DEGB analogue of the Buchdahl bound depends on both $M$ and $R$
and is less restrictive than $4/9$ for $\alphaGB > 0$; the GR lines
are shown as reference only.

For $X = 3$ [Fig.~\ref{fig:compact_pair}(a)], the compactness at the
maximum-mass configuration increases from $C = 0.2561$ ($\alphaGB = 0$)
to $C = 0.2610$ ($\alphaGB = 1\,\mathrm{km}^2$) and $C = 0.3123$
($\alphaGB = 10\,\mathrm{km}^2$), values that lie comfortably below the
GR Buchdahl bound throughout. An important feature of these curves is
that the filled circle marking $M_\text{max}$ does not coincide with
the maximum of $C$ along the sequence. The compactness continues to
increase along the dynamically unstable branch (dashed), reaching
values up to $C \approx 0.34$ for $\alphaGB = 10\,\mathrm{km}^2$.
This reflects the fact that on the unstable branch the radius decreases
faster than the mass, so the ratio $M/R$ continues to grow even as
$dM/d\rhoC < 0$. Physically, a configuration with $\rhoC$ beyond the
turning point would collapse rather than represent a stable equilibrium,
so the maximum physically realisable compactness coincides with the
maximum-mass configuration.

For $X = 4$ [Fig.~\ref{fig:compact_pair}(b)], the compactness values at
$M_\text{max}$ are $C = 0.2531$, $0.2552$, and $0.2755$ for $\alphaGB
= 0$, $1$, and $10\,\mathrm{km}^2$, respectively. These are
systematically lower than the corresponding $X = 3$ values, a direct
consequence of the larger stellar radii of the stiffer EOS. The
unstable branch again exhibits rising $C$ beyond the turning point,
with the $\alphaGB = 10\,\mathrm{km}^2$ curve reaching $C \approx 0.30$
at the highest densities shown. In both panels, all stable
configurations remain well below the GR Buchdahl bound $C = 4/9$,
confirming that the present quark star models do not violate the
compactness constraint of classical relativistic fluid-sphere theory.

\subsection{Radial Sound-Speed Profiles}
\label{subsec:cs2_profile}

The radial profiles of the squared speed of sound $\cs^2(r)$ through
the interior of the maximum-mass configuration are shown in
panels (a) and (b) of Fig.~\ref{fig:cs2_r_pair} for $X = 3$ and
$X = 4$, respectively. At each radial point $r$, the local baryon
chemical potential $\muB(r)$ is recovered from the pressure profile
via the EOS interpolator, and $\cs^2$ is evaluated from
Eq.~\eqref{eq:cs2}. Since the FKV[$X$] EOS is independent of
$\alphaGB$, the three profiles for different values of $\alphaGB$
reflect the same underlying thermodynamic relation mapped onto
different stellar structures: increasing $\alphaGB$ modifies the
metric and hence the radial pressure gradient, which shifts the
mapping $r \leftrightarrow \muB(r)$ and thereby produces the
observable separation between the curves.

For $X = 3$ [Fig.~\ref{fig:cs2_r_pair}(a)], the central values are
$\cs^2(0) = 0.2914$, $0.2917$, and $0.2958$ for $\alphaGB = 0$,
$1$, and $10\,\mathrm{km}^2$, respectively. The small but systematic
increase with $\alphaGB$ reflects the fact that higher coupling
selects a more centrally dense maximum-mass star, probing a slightly
higher $\muB$ at the centre and hence a slightly different region of
the EOS. Moving outward from the centre, all three profiles decrease
monotonically and reach a common minimum of $\cs^2 \approx 0.282$
near $r \approx 8.5\,\mathrm{km}$, beyond which the profiles
converge and rise steeply together to $\cs^2 \approx 0.315$ at the
stellar surface. The convergence near the surface is a direct
consequence of the common EOS surface condition: all configurations
share the same surface chemical potential $\muB^{(0)}(X=3) \approx
0.913\,\mathrm{GeV}$, and therefore the same $\cs^2$ at the surface
regardless of $\alphaGB$. The non-monotonic profile --- decreasing
from the centre then rising steeply in the outer layers --- reflects
the minimum of the FKV[3] sound speed in the intermediate $\muB$
range (cf.~Fig.~\ref{fig:eos_cs2}), now mapped into the stellar
interior. Crucially, $\cs^2 < 1/3$ throughout the entire stellar
interior for all three configurations, confirming that the quark
matter remains sub-conformal inside the maximum-mass star.

For $X = 4$ [Fig.~\ref{fig:cs2_r_pair}(b)], the central values are
$\cs^2(0) = 0.2868$, $0.2870$, and $0.2890$ for $\alphaGB = 0$,
$1$, and $10\,\mathrm{km}^2$, respectively, with the three curves
nearly indistinguishable throughout. The minimum is reached near
$r \approx 14\,\mathrm{km}$ at $\cs^2 \approx 0.269$, and the
subsequent rise to the surface value is considerably gentler than
for $X = 3$, reaching only $\cs^2 \approx 0.278$ at the surface.
The narrower range of $\cs^2$ variation ($\Delta\cs^2 \approx 0.02$
for $X = 4$ versus $\approx 0.035$ for $X = 3$) reflects the stiffer
EOS probing a lower and more restricted range of $\muB$ across the
stellar interior. The sub-conformal condition $\cs^2 < 1/3$ is again
satisfied pointwise throughout.

\subsection{Maximum-Mass Summary}
\label{subsec:mmax}

Figure~\ref{fig:Mmax_summary} summarises the dependence of $\Mmax$
on the Gauss--Bonnet coupling $\alphaGB$ for both values of the
renormalization-scale parameter $X$, alongside the current
observational constraints. For both $X = 3$ and $X = 4$, $\Mmax$
increases monotonically with $\alphaGB$, consistent with the
interpretation that positive higher-curvature corrections weaken
effective gravitational attraction and permit higher maximum masses.

For $X = 3$, the increase is substantial: from $2.0431\,\Msun$ at
$\alphaGB = 0$ to $2.5013\,\Msun$ at $\alphaGB = 10\,\mathrm{km}^2$,
a fractional increase of $22.4\%$. The GR baseline and the
$\alphaGB = 1\,\mathrm{km}^2$ configuration ($2.0861\,\Msun$) both
lie below the PSR~J0952$-$0607 band
($2.35 \pm 0.17\,\Msun$~\cite{Romani:2022psr}) and the GW190814
secondary mass
($2.59^{+0.08}_{-0.09}\,\Msun$~\cite{Abbott:2020gw190814}). The
$\alphaGB = 10\,\mathrm{km}^2$ configuration enters the lower portion
of the PSR~J0952$-$0607 band, falling short of the GW190814 constraint
by $\approx 0.09\,\Msun$. If the GW190814 secondary is interpreted as
a quark star, the FKV[3] EOS would require a
Gauss--Bonnet coupling beyond $10\,\mathrm{km}^2$, which lies outside
the current observational bound on $\alphaGB$.

For $X = 4$, the GR baseline already yields $\Mmax = 3.0415\,\Msun$,
well above both observational constraints. The increase from $\alphaGB
= 0$ to $10\,\mathrm{km}^2$ is $9.6\%$ in fractional terms
($3.0415 \to 3.3324\,\Msun$), smaller than for $X = 3$ owing to the
lower central densities of the stiffer EOS, where the Gauss--Bonnet
correction is less effective. The entire $X = 4$ sequence lies above
both the GW190814 and PSR~J0952$-$0607 bands at all values of
$\alphaGB$ considered. This indicates that if the FKV[4] EOS is correct, even GR quark stars would be massive enough to
account for the GW190814 secondary and PSR~J0952$-$0607 without any
modification to gravity. In this sense the $X = 4$ results constrain
not the gravity sector but rather the EOS sector: an independent
measurement of the radius of a $\sim3\,\Msun$ compact object would be
needed to distinguish between the GR and 4DEGB predictions for this
EOS.

The maximum-mass results for all six $(X, \alphaGB)$ configurations
are collected in Table~\ref{tab:results}.

\begin{table}[ht]
\centering
\caption{Maximum-mass configurations for quark stars in 4DEGB gravity
with the FKV[$X$] EOS. Listed are the renormalization-scale
parameter $X$, the Gauss--Bonnet coupling $\alphaGB$, the maximum
gravitational mass $\Mmax$, the corresponding stellar radius $R$,
the compactness $C = \Mmax/R$, and the central density $\rhoC$. All
results use $\muB^{\max} = 2.0\,\mathrm{GeV}$ (FKV validity bound).}
\label{tab:results}
\begin{tabular}{cccccc}
\toprule
$X$ & $\alphaGB$ $[\mathrm{km}^2]$ &
$\Mmax$ $[M_{\odot}]$ & $R$ $[\mathrm{km}]$ &
$C$ & $\rhoC$ $[\mathrm{g\,cm}^{-3}]$ \\
\midrule
3 & 0  & 2.0431 & 11.78 & 0.2561 & $1.68\times10^{15}$ \\
3 & 1  & 2.0861 & 11.80 & 0.2610 & $1.73\times10^{15}$ \\
3 & 10 & 2.5013 & 11.83 & 0.3123 & $2.38\times10^{15}$ \\[3pt]
4 & 0  & 3.0415 & 17.75 & 0.2531 & $7.44\times10^{14}$ \\
4 & 1  & 3.0698 & 17.77 & 0.2552 & $7.54\times10^{14}$ \\
4 & 10 & 3.3324 & 17.85 & 0.2755 & $8.57\times10^{14}$ \\
\bottomrule
\end{tabular}
\end{table}

\section{Conclusions}
\label{sec:conclusions}

In this work, we have investigated the equilibrium structure and
stability of selfbound quark stars within the framework of regularized
four-dimensional Einstein--Gauss--Bonnet (4DEGB) gravity, employing
the perturbative QCD equation of state of Fraga, Kurkela and
Vuorinen~\cite{Fraga:2014interacting}. The FKV[$X$] EOS is
characterised by a single renormalization-scale parameter $X$, contains
no bag constant, and defines the stellar surface entirely through the
vanishing of the quark-matter pressure. We solved the modified
Tolman--Oppenheimer--Volkoff equations derived from the scalar--tensor
formulation of 4DEGB gravity~\cite{Hennigar:2020lsl,Gammon:2023uss}
for two representative values $X \in \{3, 4\}$ and three values of the
Gauss--Bonnet coupling $\alphaGB \in \{0, 1, 10\}\,\mathrm{km}^2$,
scanning 300 central density values per $(X, \alphaGB)$ cell with a
GR-recovery validation at the $0.1\%$ level.

Our main findings regarding the mass--radius relations are as follows.
For the soft EOS ($X = 3$), the GR baseline yields a maximum mass of
$\Mmax = 2.0431\,\Msun$, which increases to $2.0861\,\Msun$ at
$\alphaGB = 1\,\mathrm{km}^2$ and to $2.5013\,\Msun$ at
$\alphaGB = 10\,\mathrm{km}^2$, a fractional enhancement of $22.4\%$.
Only the $\alphaGB = 10\,\mathrm{km}^2$ configuration enters the
PSR~J0952$-$0607 mass band ($2.35 \pm 0.17\,\Msun$~\cite{Romani:2022psr}),
while the GW190814 secondary
($2.59^{+0.08}_{-0.09}\,\Msun$~\cite{Abbott:2020gw190814}) remains
beyond reach for the entire $X = 3$ branch within the current
observational bound on $\alphaGB$. For the stiff EOS ($X = 4$), the
GR baseline already yields $\Mmax = 3.0415\,\Msun$, with a further
increase to $3.3324\,\Msun$ at $\alphaGB = 10\,\mathrm{km}^2$. The
entire $X = 4$ sequence lies above both observational constraints at
all values of $\alphaGB$ considered. These results demonstrate that
the higher-curvature corrections systematically enhance the maximum
supported mass, while the EOS stiffness controls whether observational
thresholds are exceeded already at the GR level.

The compactness analysis confirms that all stable configurations
remain well below the GR Buchdahl bound $C = 4/9$ throughout.
The compactness at $M_\text{max}$ ranges from $C = 0.2531$ ($X=4$,
GR) to $C = 0.3123$ ($X=3$, $\alphaGB = 10\,\mathrm{km}^2$). An
important structural feature of the compactness sequences is that
the maximum-mass turning point does not coincide with the maximum
compactness along the sequence: $C$ continues to rise on the
dynamically unstable branch, reaching values up to $C \approx 0.34$
for $X = 3$ at $\alphaGB = 10\,\mathrm{km}^2$. The 4DEGB Buchdahl
bound is less restrictive than its GR counterpart for $\alphaGB > 0$,
and all configurations satisfy it with a comfortable margin. The
mass--central-density sequences confirm that increasing $\alphaGB$
shifts the turning point to higher central densities, with $\rhoC$
at $M_\text{max}$ ranging from $7.44\times10^{14}\,\mathrm{g\,cm}^{-3}$
($X=4$, GR) to $2.38\times10^{15}\,\mathrm{g\,cm}^{-3}$ ($X=3$,
$\alphaGB = 10\,\mathrm{km}^2$), consistent with the stabilising role
of positive higher-curvature corrections.

The radial sound-speed profiles through the interior of the
maximum-mass configurations confirm that $\cs^2 < 1/3$ holds
pointwise throughout the stellar interior for all $(X, \alphaGB)$
cells, establishing that the FKV[$X$] quark matter is everywhere
sub-conformal inside the star. The profiles exhibit a non-monotonic
radial structure: $\cs^2$ decreases from the central value outward,
reaches a minimum near $r \approx 8.5\,\mathrm{km}$ ($X = 3$) or
$r \approx 14\,\mathrm{km}$ ($X = 4$), and then rises steeply toward
the stellar surface. This behaviour reflects the minimum of the
FKV[$X$] sound speed at intermediate $\muB$, now mapped into the
stellar interior through the radial pressure gradient. The
Gauss--Bonnet coupling $\alphaGB$ enters the profiles only through
this mapping: the central $\cs^2$ increases slightly with $\alphaGB$
as higher coupling selects a more compact and denser maximum-mass
configuration, while all three curves converge to the same surface
value determined by the common surface chemical potential
$\muB^{(0)}(X)$.

Several limitations of the present analysis should be noted. The
FKV[$X$] pocket formula is valid for $\muB \lesssim 2\,\mathrm{GeV}$
and $X \in [1,4]$; results for the softer $X = 3$ EOS approach the
lower end of this range at the stellar surface. The present
calculation is restricted to non-rotating stellar models; rotation
can modify the maximum mass and compactness substantially for
millisecond pulsars such as PSR~J0952$-$0607. Tidal deformability
and gravitational-wave signatures have not been computed here and
would be necessary for a direct comparison with binary merger
observations. Finally, the $\alphaGB = 10\,\mathrm{km}^2$
configuration sits at the current observational bound on the coupling,
so results at this value should be interpreted as representing the
maximal allowed deviation from GR rather than a preferred model.

Future investigations should extend the present work in several
directions. The computation of tidal deformability parameters $\Lambda$
for the FKV[$X$] EOS in 4DEGB gravity would enable direct comparison
with the GW170817 tidal constraints and place joint bounds on $(X,
\alphaGB)$. The incorporation of rotation through the Hartle--Thorne
formalism would allow a more complete assessment of the PSR~J0952$-$0607
constraint. Extension to other modified gravity frameworks --- such as
dRGT massive gravity, scalar--tensor theories, and $f(R)$ gravity ---
with the same pQCD EOS would provide a systematic comparison of
higher-curvature effects on quark star observables. Finally, the use
of perturbative QCD results beyond the three-loop order underlying
the FKV[$X$] pocket formula, including the recently computed
N$^3$LO contributions~\cite{Gorda:2018nexttonexttonexttoleading,
Gorda:2021cold}, would improve the theoretical accuracy of the EOS
and reduce the renormalization-scale uncertainty encoded in $X$.

\section*{Declaration of competing interest}
The authors declare that there are no financial or personal relationships that could have influenced the research
presented in this paper.
\section*{ Data availability}
No data was used for the research described in the article.

\section*{Acknowledgements}

This work was supported by the Deanship of Scientific Research, Vice Presidency for Graduate Studies and Scientific Research, King Faisal University, Saudi Arabia (Grant No: KFU263249).

\bibliographystyle{apsrev4-1}
\bibliography{master_refs}

\end{document}